# Intracranial hemodynamics simulations: An efficient and accurate immersed boundary scheme

D. S. Lampropoulos[1], G. C. Bourantas[2], B. F. Zwick[2], G. C. Kagadis[3,4], A. Wittek[2], K. Miller[2,5], V. C. Loukopoulos[1]

[1]*Department of Physics, University of Patras, Patras, 26500, Rion, Greece*
[2]*Intelligent Systems for Medicine Laboratory, The University of Western Australia, 35 Stirling Highway, Perth, WA 6009, Australia*
[3]*Department of Medical Physics, School of Medicine, University of Patras, Rion, GR 26504, Greece*
[4]*Department of Imaging Physics, The University of Texas MD Anderson Cancer Center, Houston, TX 77030, USA*
[5]*Harvard Medical School, 25 Shattuck St, Boston, MA 02115, USA*

**Abstract:** Computational fluid dynamics (CFD) studies have been increasingly used for blood flow simulations in intracranial aneurysms (ICAs). However, despite the continuous progress of body-fitted CFD solvers, generating a "high quality" mesh is still the bottleneck of the CFD simulation, and strongly affects the accuracy of the numerical solution. To overcome this challenge, which will allow preforming CFD simulations efficiently for a large number of aneurysm cases we use an Immersed Boundary (IB) method. The proposed scheme relies on Cartesian grids to solve the incompressible Navier-Stokes (N-S) equations, using a finite element solver, and Lagrangian points to discretize the immersed object. All grid generations are conducted through automated algorithms which require no user input. Consequently, we verify the proposed method by comparing our numerical findings (velocity values) with published experimental results. Finally, we test the ability of the scheme to efficiently handle hemodynamic simulations on complex geometries on a sample of four patient-specific intracranial aneurysms.

## 1. Introduction

*1.1. Blood flow*

Intracranial aneurysms (ICAs) are pathological dilations of the cerebral arteries, which when ruptured can lead to subarachnoid haemorrhage (SAH) and stroke [1]. ICAs are usually located at bifurcation points, inside or near the Circle of Willis. Unruptured ICAs are often diagnosed accidentally [2]. However, treatment of ICAs is well established, and surgical clipping or endovascular treatment are routinely applied [3, 4]. Nevertheless, since surgical treatment carries inherent risks, the ideal situation would be to detect and treat only those aneurysms that are likely to rupture. The assessment of patient-specific rupture risk for ICAs remains one of the most challenging medical problems for neuro-radiologists and neurosurgeons, since the mechanisms initiation, progression and, finally, rupture of an aneurysm are not yet completely understood. It is therefore important to develop computational tools, which will provide reliable information on the local hemodynamics.

Computational fluid dynamics (CFD) simulations [5-7] are frequently used to study the initiation and the progression of ICAs, by correlating the calculated hemodynamic parameters with the biological phenomena such as Wall Shear Stress (WSS) that occur on the arterial wall [8]. The majority of CFD models developed for ICAs hemodynamics utilize body-fitted methods such as finite volume (FV) or finite element (FE) method [6, 7, 9, 10]. In all the mesh-based methods mesh quality strongly affects the accuracy of the numerical solution, and despite that mesh generators have evolved over the last years, generating high-quality meshes is not trivial and often leads to time-consuming human involvement. Thus, despite some efforts [11], the mesh-based methods still struggle to efficiently scale to large sample sizes as there is an absence of standardization protocols [12] and automated pipe-lined procedures where limited or no human intervention is required [5]. In addition to all the aforementioned aspects, the huge variation in CFD methodologies used, often leads to questionable results [13, 14] and prevents the usage of CFD simulations as a clinical tool.

*1.2 Immersed boundary methods*

Immersed boundary (IB) method-based models are an attractive alternative to the mesh-based numerical methods, to model flows in complex geometries, stationary and/or moving objects. In contrast to mesh-based methods, IB methods avoid tedious mesh generation (they use a Cartesian grid) to solve the flow equations and discrete Lagrangian points for the implementation of no-slip boundary condition [15].

IB method has been introduced by Peskin [16] for modeling blood flow through a beating heart. Flow equations are solved on a uniform Cartesian grid (Eulerian nodes), while the boundary of the solid is described through a set of nodes (Lagrangian nodes) which, in general, do not coincide with the fluid grid points. Fluid and solid nodes communicate through volume forces, which are used to enforce the velocity boundary condition and ensure the conservation of linear momentum. Several variants of Peskin's original immersed boundary method have been developed and applied to various physical problems [15]. The existing IB methods are divided in two major groups. In the first group of techniques (widely used in problems involving sharp boundaries and rigid objects), referred as "continuous forcing" body forces are derived prior to the discretization step [17, 18]. The second group, termed as "discrete forcing" methods is based on a set of singular body forces, defined on the Eulerian fluid nodes, to enforce the desired boundary values [19, 20].

IB methods are quite popular for external flow simulations [15, 21, 22]. Although, all the advantageous features of IB methods, which made them popular for external flow simulations, do not entirely apply for internal flows. In fact their use on internal biological flows is still limited [23]. The advantageous features of Cartesian grid flow solvers, when applied to internal flows lead to issues related to their efficiency. For internal flows with complex geometries, only a subset of the total grid points is located inside the fluid domain, while the remaining points fall outside. The external grid points are not used in the numerical solution and increase the computational cost. In fact, in many cases the external grid points are a large percentage of the total grid points, which is the most profound cause for the computational inefficiency of the IB method on external flow cases. Anupindi et al. [23] proposed a multiblock based IB method for anatomical geometries. They reduced the number of grid elements in the exterior of the fluid domain while retaining the effectiveness of simple Cartesian grids. Dillard et al. [24] proposed an alternative framework with an image-to-computation algorithm designed for minimum user intervention. To further eliminate time-consuming human interventions, Seo et al. [25] proposed a highly automated computational procedure utilizing the sharp-interface immersed boundary method. Spacing for the Cartesian grid for the flow simulation was extracted from the voxel spacing of the angiogram data.

*1.3 Contributions of this study*

In the this study, we examine the applicability of the previously developed and validated finite element (FE) [26] boundary condition enforced immersed boundary method [27], to simulate blood flow in patient-specific ICAs. We utilize Cartesian Eulerian grids for the solution of the flow field and Lagrangian points to represent the immersed object. We maximize the efficiency of the present scheme by utilizing local mesh refinement. The simulation pipeline (solution procedure) is user friendly, since both Eulerian mesh and Lagrangian points are generated automatically. We demonstrate the reproducibility and the versatility of the presented scheme simulating blood flow on a sample of four patient-specific intracranial aneurysms. A naïve implementation of the fluid flow solver used in this study is available on GitHub (https://github.com/DmLabrop/Oasis_IMB).

The remainder of the paper is organized as follows. In Sections 2 and 3, we present the numerical methodology and the benchmark cases used for verification, respectively. Section 4 is concerned with a numerical example that demonstrates the accuracy and applicability of the proposed IB scheme. Finally, Section 5, contains the conclusions of our study.

## 2. Numerical method

### 2.1 Governing equations

In this section, we briefly describe the boundary condition-enforced immersed boundary finite element solver, used in the numerical solution of the incompressible Navier-Stokes equations. We consider a rectangular (or square) computational domain (Eulerian domain) $\Omega$, which involves an immersed body described in the form of a closed surface $\partial\Omega^L$. The immersed object is modelled as localized body forces acting on the surrounding fluid. Therefore, the IB formulation for the incompressible Navier-Stokes equations involving immersed objects is expressed in the primitive variables (velocity $u$ and pressure $p$) formulation as

$$\frac{\partial \boldsymbol{u}}{\partial t} + (\boldsymbol{u} \cdot \nabla)\boldsymbol{u} = v_f \nabla^2 \boldsymbol{u} - \nabla p + \boldsymbol{f} \tag{1}$$

$$\nabla \cdot \boldsymbol{u} = 0 \tag{2}$$

subject to the no-slip boundary condition on $\partial\Omega^L$

$$\boldsymbol{u}(\boldsymbol{X}(s), t) = \boldsymbol{U}_B \tag{3}$$

with $p$ being the fluid pressure, $\boldsymbol{u}$ the fluid velocity, $\boldsymbol{U}_B$ the prescribed velocity on the immersed boundary, and $v_f$ the kinematic viscosity of the fluids. The external force $\boldsymbol{f}$, which represents the interaction force between the fluid and the immersed object (immersed boundary), is given as

$$f(x,t) = \int F(x,t)\delta(x - X(s,t))ds \tag{4}$$

where $\delta(x - x_i)$ is the Dirac delta function, $X_i$ is the position of the Lagrangian points (located on the immersed boundary), $x$ is the position of the Eulerian grid, and $F_i(X_i)$ is the force exerted on the Lagrangian point $X_i$. Additionally, the velocity at the immersed boundary nodes is interpolated from the velocity at the Eulerian nodes as

$$u(x - X(s,t)) = \int u(x)\delta(x - X(s,t))dV. \tag{5}$$

*2.2 IB method algorithmic procedure*

The IB methodology used in this study, has been discussed in detail in a recent publication by Bourantas et al. [27], demonstrating the accuracy of the IB method. The IB combines the FE and the BCE-IB methods for internal flows, emphasizing in blood flow simulations. Herein, we replace the incremental pressure correction scheme (IPCS) flow solver used in Bourantas et *a*l. [27] (IPCS is a modified version of projection method, which provides improved accuracy at little extra computational cost), to the verified [26] and validated [28] finite element CFD solver Oasis, which is a high fidelity finite element Navier-Stokes solver using building blocks from the FEniCS project [29]. In Oasis, special care has been taken to ensure a kinetic energy-preserving and minimally-dissipative numerical solution of the Navier-Stokes equation. To avoid time step limitations, we utilize an implicit Adams–Bashforth and a semi-implicit Crank–Nicholson scheme to discretize the convective and the viscous term, respectively. Both discretization schemes are second-order accurate in time. More details considering the Oasis solver can be found in Mortensen et al. [26].

We use linear Lagrange elements ($P_1$/$P_1$) for both velocity and pressure. The solver uses unstructured meshes to discretise the flow domain and is particularly fitted to large-scale applications in complex geometries on massively parallel clusters. The main advantage of the BCE-IB method is that it satisfies accurately both the governing equations and boundary conditions using velocity and pressure correction procedures. The velocity correction applies implicitly such that the velocity on the immersed boundary (Lagrangian points) interpolated from the corrected velocity values computed on the mesh nodes (Eulerian nodes) accurately satisfies the prescribed velocity boundary conditions.

*2.2 Solution procedure*

The procedure regarding the setup of the proposed scheme, presented in the present study, is outlined below:

1. We use 3D models derived from medical image data (DICOM) as the starting point. We generate a body-fitted tetrahedral mesh which is referred as visualization mesh. The visualization mesh is not part of the solution method as it only used to visualize the flow fields.
2. We create a rectangular domain, discretised using linear tetrahedral elements (Cartesian grid) through a simplified and automated algorithm which encloses the immersed geometry. The nodal Cartesian grid spacing $h$ is set by the user and it is the only manual operation at this stage.
3. Boundaries on the Cartesian grid such inlet, outlets, and wall are set through an automated algorithm which requires no user input.
4. We set the inflow and outflow conditions.
5. We refine visualization mesh to obtain the immersed geometry nodes (Lagrangian points). The automated refinement procedure requires no user intervention and includes the creation of a surface tetrahedral mesh, where the area of the $i^{th}$ surface element $\Delta S_i$ is approximately equal to $h^2$, with $h$ being the Cartesian grid spacing.
6. We perform flow simulations on the Cartesian grid.
7. To visualize the computed velocities and to calculate velocity derived indices like Wall Shear Stress (WSS), we interpolate the solution from the Cartesian grid to the visualization mesh. To interpolate the computed solution to the visualization mesh we use FeniCS [29] build-in functions. No user interventions are required in this step.

## 3. Verification

We verify the accuracy of the proposed numerical scheme by comparing our numerical findings with the experimental data on the three-dimensional 90° degrees angle tube bend [30]. The U-bend geometry is embedded within a box domain with dimensions $-0.0064\ m \leq x \leq 0.032\ m$, $0\ m \leq y \leq 0.0384\ m$ and $-0.0064\ m \leq x \leq 0.0064\ m$ as shown in Fig. 1. We discretise the spatial/slow box domain using a high-quality tetrahedral mesh, generated using a uniform Cartesian grid (to generate the mesh we used the FEniCS build-in function BoxMesh). The tetrahedral mesh is locally refined in the vicinity and within the interior of the immersed boundary (U-bend tube geometry), using the FEniCS built-in function Refine [29] which uses the algorithm by Plaza and Carey [31].

We solve the flow equations (velocity and pressure) on a Cartesian mesh of tetrahedral linear elements ($P_1/P_1$). The inner diameter $D_i$ and the curvature radius $R$ of the tube were set to 4 and 24 mm, respectively. We use a parabolic velocity profile ($u_z(r) = U_{max}\left(1 - \left(\frac{r}{R}\right)^2\right)$) at the inlet (green circle in Fig. 1), and zero pressure ($p = 0$) boundary conditions at the outlet, while no-slip boundary conditions are set at the remaining surfaces faces of the computational domain, as shown in Fig. 1. The Reynolds number is $Re = 300$. To reduce flow disturbances due to inflow boundary conditions, we extend the inlet and outlet by one diameter $R_D$ in length (4 mm). Axial velocities at various angles close to the tube bend were measured using laser Doppler velocimetry [30]. The grid is locally refined close and in the enclosed region of the immersed object, as shown in Figure 1 increasing the percentage of the elements inside the immersed object up to 85%. The resulting Cartesian grid consists of 2,942,170 tetrahedral elements (Fig. 1) with a minimum grid spacing settling to $2.0 \times 10^{-4} m$.

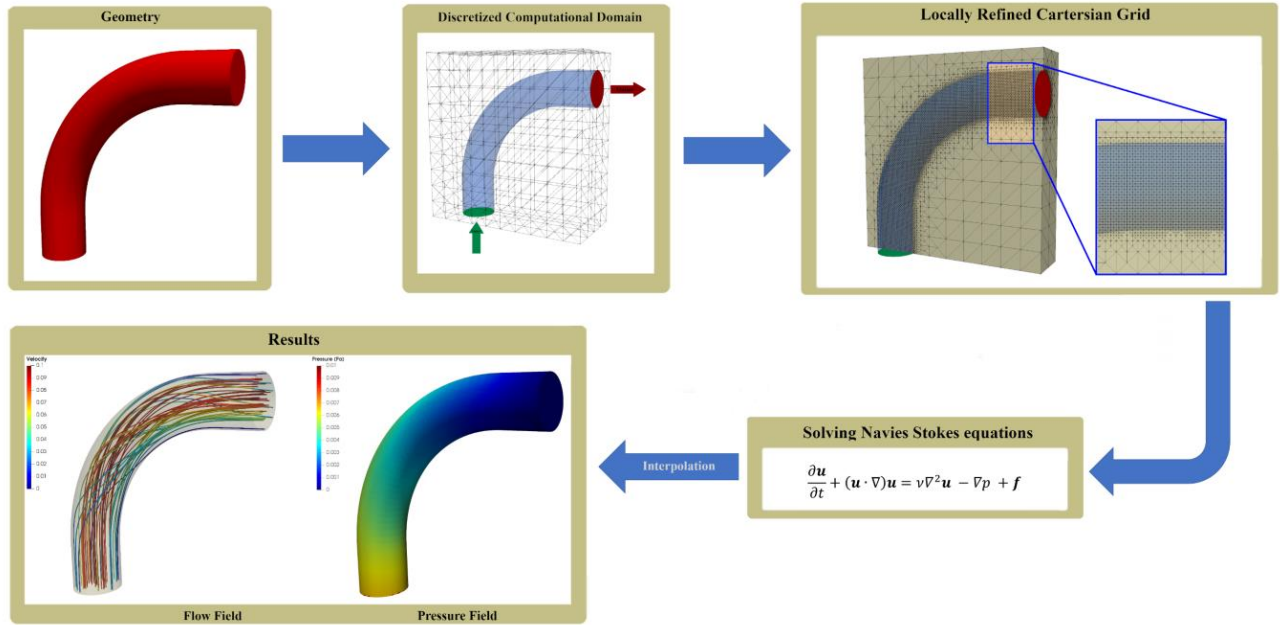

**Figure 1.** The solution workflow for the 90° degrees angle tube bend case.

The numerical solution obtained from the Cartesian grid, is interpolated on a tetrahedral visualization mesh (see Section 2.2 Step 7). Figure 2 shows the computed axial velocity plotted along with the experimental data [30], measured at the central cross-section at the end of the tube's curved section (tube outlet). The numerical results obtained using the linear ($P_1/P_1$) elements are in good agreement the experimental data, highlighting the accuracy of the proposed scheme [27].

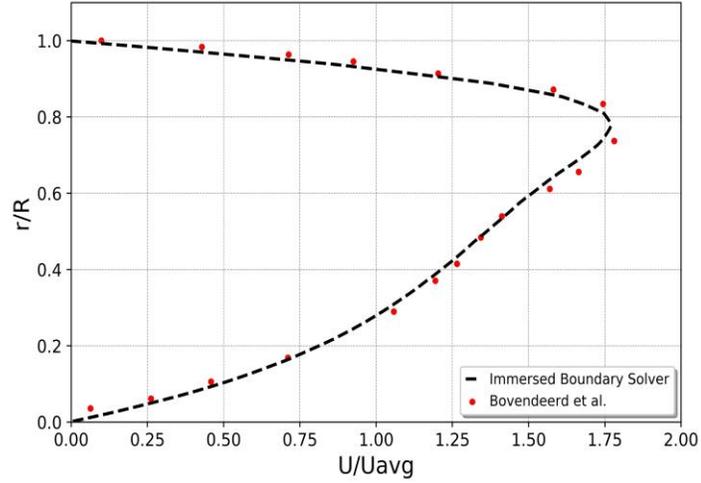

**Figure 2.** Axial velocity (dashed line) using and the experimental data (red dots) at the tube outlet.

## 4. Numerical results

In this section we demonstrate the applicability of the proposed IB method on patient-specific geometries of intracranial aneurysms (ICAs). The aneurysms are located at the anterior cerebral artery (ACA), the basilar artery (BAS), the internal carotid artery (ICA) and the middle cerebral artery (MCA), as shown Fig. 3. We simulate blood flow in the aneurysmal geometries to extract the flow characteristics.

**Table 1.** Geometrical characteristics of the aneurysm geometries used in the present study.

| *Aneurisk Id* | Location | Aneurysmal dome volume (mm$^3$) | Neck Vessel angle | Size ratio | Aspect ratio | Inlet diameter (*mm*) |
|---|---|---|---|---|---|---|
| C0061 | ACA | 40.15 | 20.37 | 4.169 | 2.312 | 1.33 |
| C0068 | BAS | 86.22 | 5.98 | 2.576 | 0.897 | 1.71 |
| C0088a | ICA | 78.14 | 90.73 | 2.755 | 1.342 | 3.23 |
| C0088b | ICA | 289.19 | 56.89 | 1.922 | 1.618 | 3.36 |
| C0092 | MCA | 59.27 | 18.02 | 3.196 | 1.853 | 2.04 |

The ICAs geometries were obtained from the Aneurisk Web (http://ecm2.mathcs.emory.edu/aneuriskweb/index) online dataset repository (publicly available under the 'Creative Commons Attribution Non-Commercial 3.0 Unported License')[1]. Geometrical characteristics of the aneurysms used, such as the volume of the aneurysmal dome (mm$^3$), the angle between the neck of the aneurysm and the parent vessel, the size ratio (aneurysm to vessel size ratio), the aspect ratio (quotient of height to width of the aneurysm neck), the diameter of the arterial lumen (mm) along with general information for the patient like age and gender, are listed in Table 1.

Accurate and efficient blood flow simulation in the cerebral vascular tree is extremely important as it can potentially give important new information to be used for evaluating patients with cerebral vascular disease. With the selection of the specific aneurysm cases we aim to demonstrate the ability of the proposed scheme to cope with complex geometries from all the possible aneurysm locations in the circle of Willis.

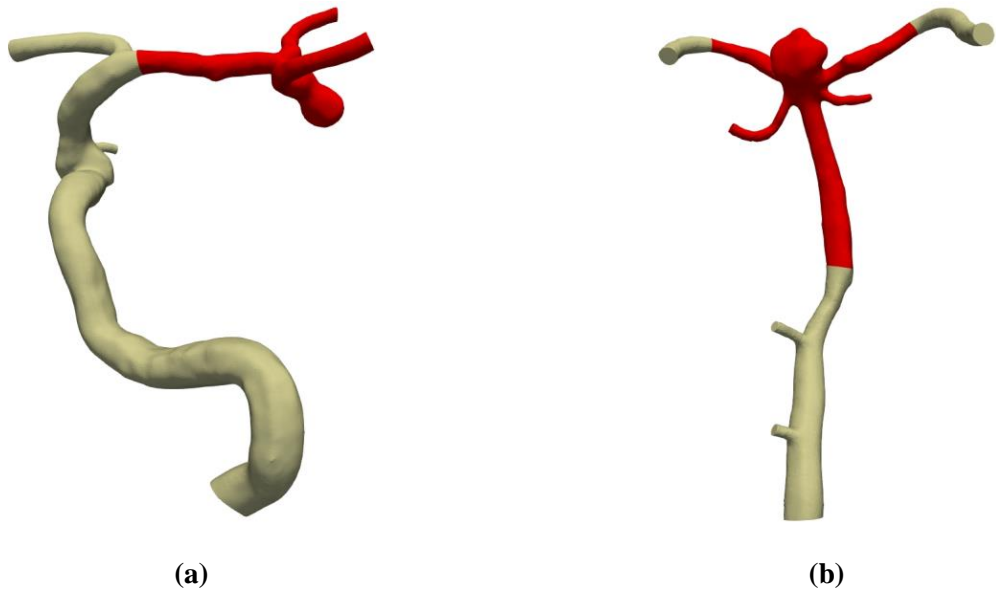

**(a)**         **(b)**

---

[1] The use of data in the Aneurisk repository has been authorized by the Ethical Committee of the Ca' Granda Niguarda Hospital

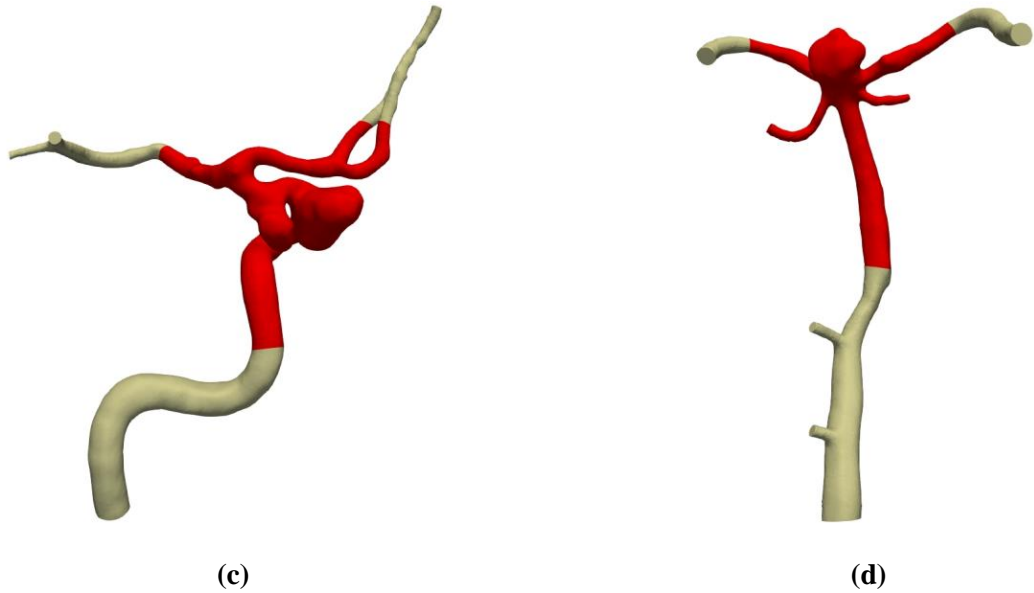

**Figure 3.** Original geometries (STL files) obtained from the Aneurisk dataset; **(a)** middle cerebral artery (MCA); **(b)** anterior communicating artery (ACA); **(c)** internal carotid artery (ICA); and **(d)** basilar artery (BAS). Highlighted with red colour are the aneurysmal geometries we used in our simulations.

In all the flow examples considered in this study, we apply the scaled (using a power-law estimation for cycle-average flow rates [32]) velocity waveform shown in Fig. 4 [33]. A fully developed Womersley velocity profile is imposed at the inlet [34], while at the outlet(s) we set pressure $p = 0$. In all the flow example considered, we simulate three cardiac cycles with a temporal resolution of $dt = 2 \times 10^{-4}$ s, which results to 4,425 time-steps per cardiac cycle [35]. We present the numerical results obtained during the third cardiac cycle, to wash out initial transients. Blood is treated as a Newtonian fluid with dynamic viscosity of $\mu = 0.00345 \; Pa \cdot s$ and density of $\rho = 1,056 \; kg/m^3$. The arterial wall is modelled as rigid [36]. To sufficiently address time-dependent flow and shear phenomena, a nodal grid spacing of at least $h_m = 1 \times 10^{-4} m$ is recommended [37]. All simulations were performed on a workstation equipped with Intel Xeon™ E5-2680 v2 processor and 32 GB of RAM.

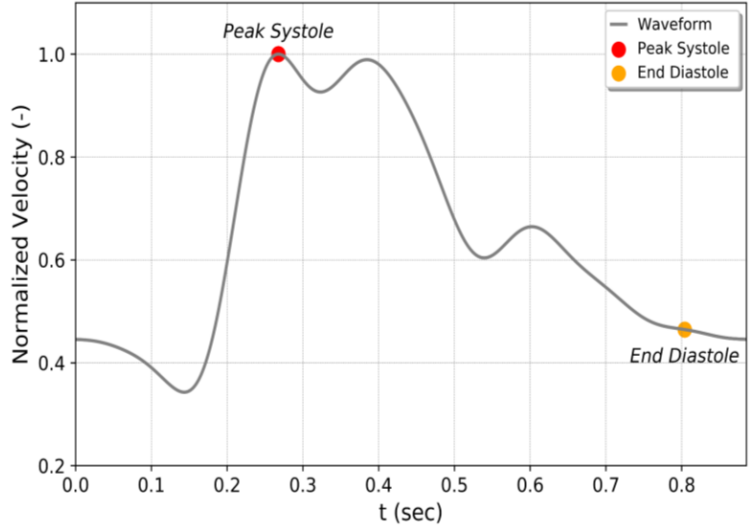

**Figure 4.** Pulsatile velocity waveform imposed at the inlet for the ICAs flow cases.

*4.1 Wall Shear Stress Calculation*

One of the most important hemodynamic index is the WSS [6, 38-40], given as

$$WSS = \|t - (t \cdot n)n\| \qquad (6)$$

with $t = \tau \cdot n$ and $\tau = 2\mu D$ being the shear stress, $D$ being the rate of deformation tensor, and $\mu$ the viscosity of the fluid. To obtain WSS we calculate the velocity spatial derivatives on the lumen wall using the modified moving least squares (MMLS) method [41]. Since we do not use patient specific inflow and outflow boundary conditions, the WSS magnitude may be slightly off the physiological range. Thus, we pay attention to the overall distribution of the WSS in every case.

*4.2 Case 1: Middle Cerebral Artery (MCA) aneurysm*

We simulate blood flow in the middle cerebral artery aneurysm case C0092 (Fig. 3a). The flow domain is a box domain. The immersed boundary (Middle Cerebral Artery aneurysm) is immersed within the box domain. In this flow example, we demonstrate the accuracy and applicability of the IB method. We conduct a grid independent study, using four successively denser grids. All grids are locally refined close and inside the immersed boundary (Fig. 5), to increase the number of the elements inside the immersed geometry (the percentage of elements inside the immersed boundary is 85%).

We generate the Cartesian linear tetrahedral element ($P_1/P_1$) grids through the automated algorithm as described in solution procedure subsection. The generated grids require no quality checks. Although grid generation times are on par with the more established body-fitted mesh generators, in our case the user input is limited to a single value (Cartesian grid nodal spacing) leading to a more user-friendly pipe-lined procedure. We interpolate the velocity values computed on Cartesian grids to the visualization volume mesh derived from the initial surface mesh, to examine the convergence of the method. We calculate the peak systolic velocity magnitude maximum absolute error defined as $L_\infty = |u_i^{coarse} - u_i^{dense}|$ and the Root Mean Square Error (RMSE) defined as $L_2 = \frac{1}{N}\sqrt{\sum_{i=1}^{N}(u_i^{coarse} - u_i^{dense})^2}$, where $u_i^{coarse}$ and $u_i^{dense}$ are the velocities of the coarse and the dense grids, respectively. The results are computed using the grid with the highest resolution as the reference solution. Table 2 contains the velocity magnitude errors between each mesh size between the four grid spacings with information regarding the generated grids.

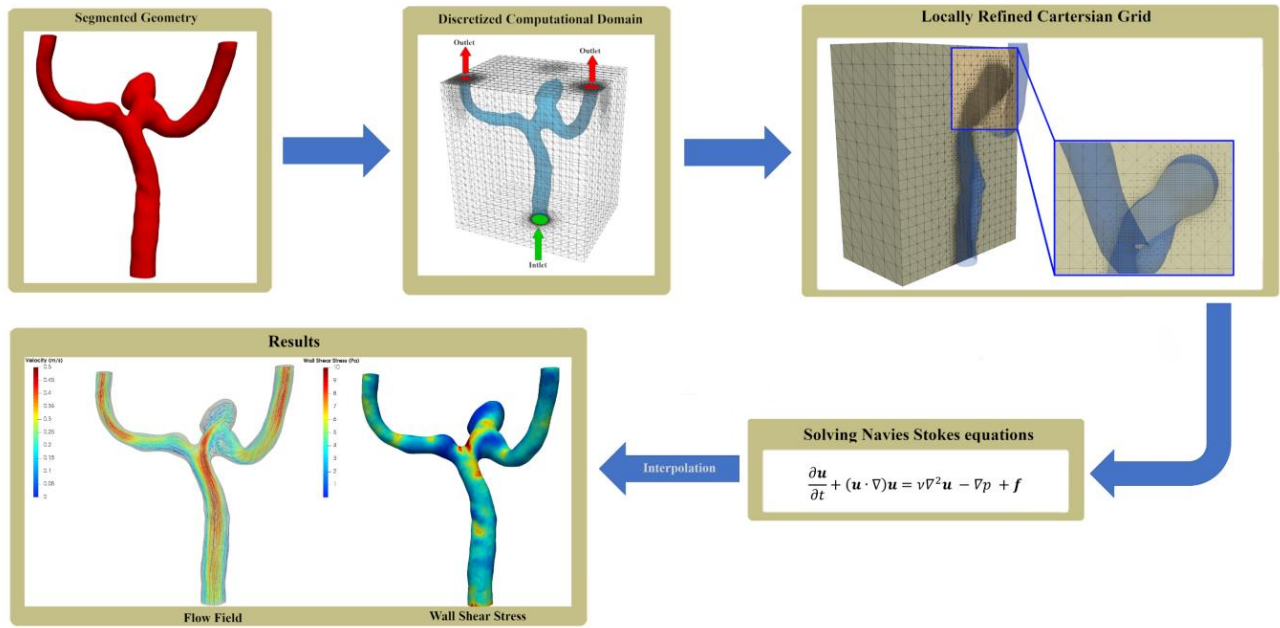

**Figure 5.** The solution workflow for the MCA aneurysm case.

**Table 2.** $L_\infty$ and $L_2$ error norms for velocity for case C0092 using the grid with highest resolution as reference.

| Minimum tetrahedral mesh nodal spacing ($m$) | $L_\infty$ | $L_2$ | Tetrahedral elements |
|---|---|---|---|
| $1.5 \times 10^{-4}$ | $8.32 \times 10^{-4}$ | $7.33 \times 10^{-3}$ | 1,997,246 |
| $1.25 \times 10^{-4}$ | $2.09 \times 10^{-4}$ | $1.44 \times 10^{-3}$ | 2,723,267 |
| $1.0 \times 10^{-4}$ | $1.37 \times 10^{-5}$ | $1.28 \times 10^{-4}$ | 3,804,084 |
| $7.5 \times 10^{-5}$ | — | — | 4,794,854 |

Both error norms show a good agreement between the coarser and finer grid. To further demonstrate the convergence rate of the proposed scheme, we set a line profile in the aneurysmal dome, where flow instabilities are generated (Fig. 6a), and we compute the velocity (magnitude) profile (Fig. 6b). Minor differences in the velocity magnitude were computed for the velocity values from the different tetrahedral meshes, highlighting the convergence of the flow field values. Both errors show a strong convergence for the computed flow field between the different Cartesian grid resolutions.

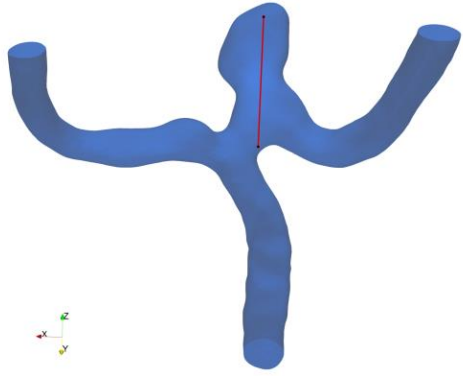 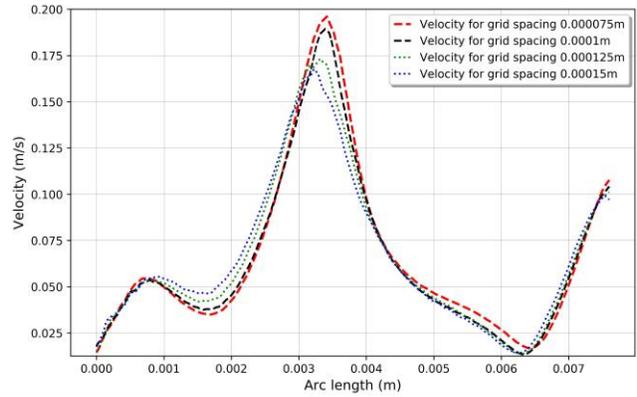

(a)                                        (b)

**Figure 6. (a)** The velocity profiles measured along the red line located inside the aneurysmal dome, and **(b)** velocity magnitude profiles for different grid resolution.

Following, we present the flow field using the denser grid which has a nodal spacing $7.5 \times 10^{-5} m$. Fig. 7 illustrates the streamlines, velocity contours and WSS magnitude contours at the peak systole pressure for the MCA case aneurysm. Streamlines are smooth along the vessel body and become more complex inside the aneurysm. Inside the aneurysmal dome we have the formation of a small impinging jet leading to a more disturbed flow leading to the formation of a small re-circulation zone. Low WSS values are located on the aneurysmal dome, while high values are located on the aneurysmal neck and on the walls of the parental artery. Low WSS occurs on the aneurysmal dome, while high on the aneurysmal neck and at less extent on the walls of the parental artery. Low WSS occurs mainly due the weak flow inside the aneurysmal dome of bifurcation aneurysms. That observation also agrees with the findings of Shojima et al. [9] and Valencia et al. [7].

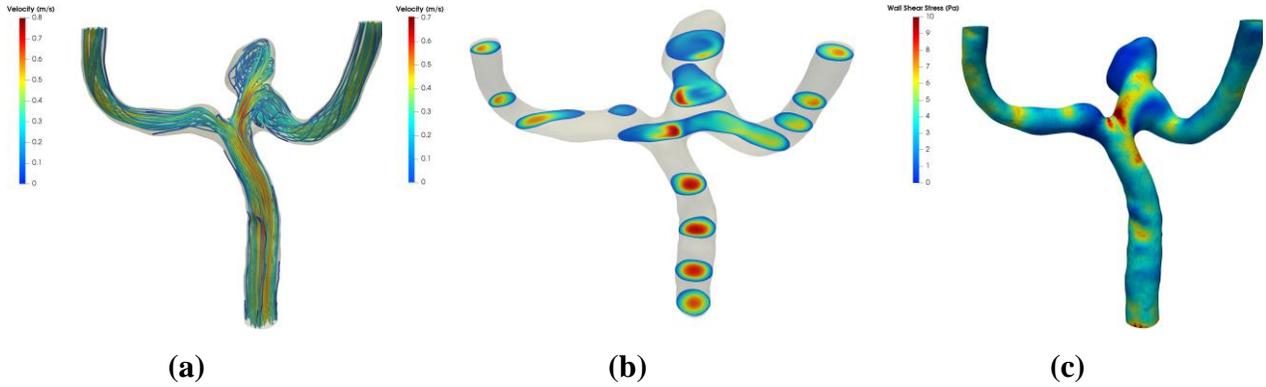

**Figure 7.** (a) Streamlines **(b)** velocity contours and **(c)** wall shear stress magnitude distribution (units in Pa) for blood flow through the MCA aneurysm example.

*4.3 Blood flow in anterior cerebral artery (ACA), internal cerebral artery (ICA) and basilar artery (BA) aneurysms*

Following, we simulate blood flow in the anterior cerebral artery (ACA) (Fig. 3b). The flow domain is a box domain, discretised using linear tetrahedral elements (3,850,106 tetrahedral elements with minimum grid spacing of resolution of $h_m = 6 \times 10^{-5} m$ (see Fig. 8)), where the aneurysm geometry is immersed within. To enhance the visualization of the results and to compute WSS we interpolate the computed results to a visualization mesh as we mentioned in the solution procedure subsection.

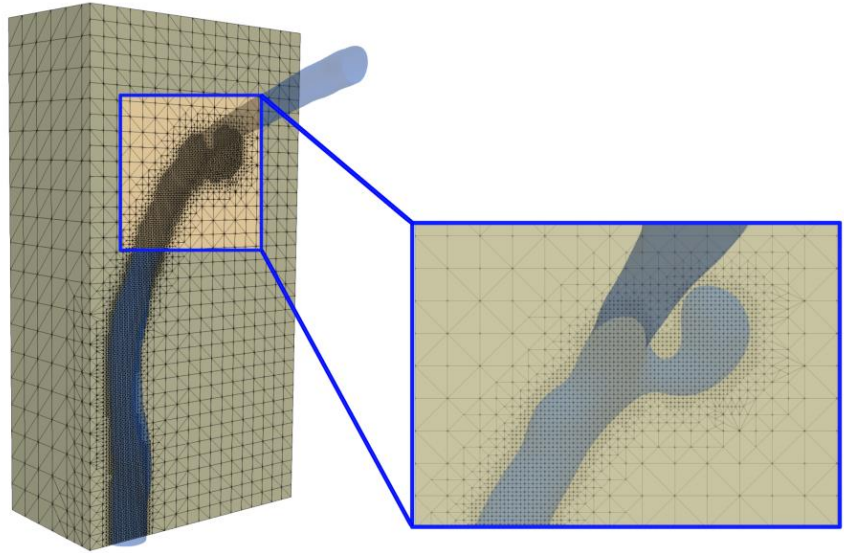

**Figure 8.** Anterior cerebral artery case Cartesian grid with four levels of refinement.

Fig. 9a shows the streamlines while Fig. 9c shows the WSS magnitude distribution at the peak systole pressure [42]. Streamlines in the parent artery align to the centreline, while in the aneurysmal dome a recirculation area is formed with more complex flow patterns, mainly due the small between the parent artery and the aneurysmal neck. Low WSS occurs on the aneurysmal dome, while high on the aneurysmal neck and at less extent on the walls of the parental artery which agrees with the findings from previous studies [6, 38].

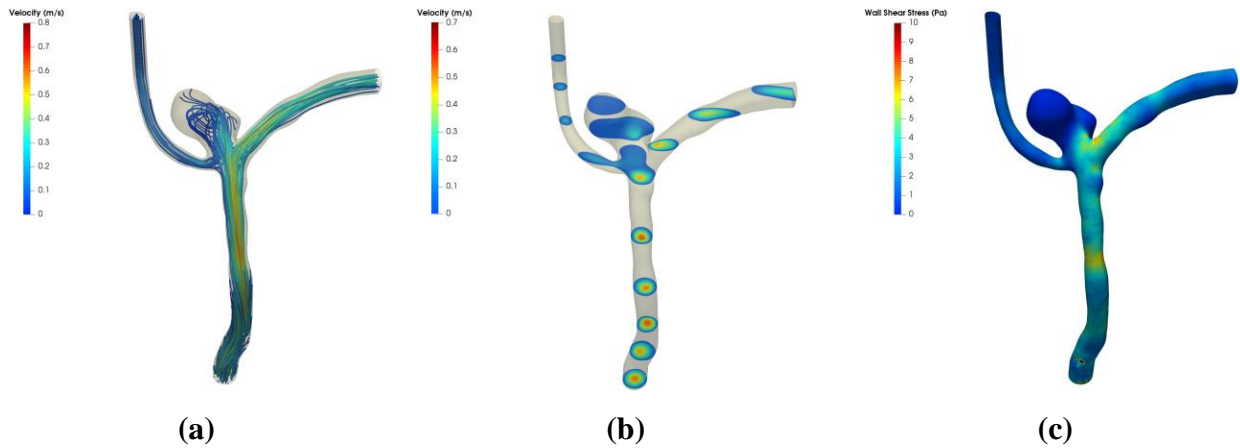

**Figure 9.** **(a)** Streamlines **(b)** velocity contours and **(c)** wall shear stress magnitude distribution (units in Pa) for blood flow through the ACA aneurysm example.

Next, we simulate blood flow in the internal cerebral artery (ICA) having two sidewall aneurysms with different volumes (Fig. 3c). The flow domain is a box domain, discretised using linear tetrahedral elements (5,818,970 tetrahedral elements with minimum grid spacing of resolution of $h = 8 \times 10^{-5} m$ (see Fig. 10)), where the aneurysm geometry is immersed within.

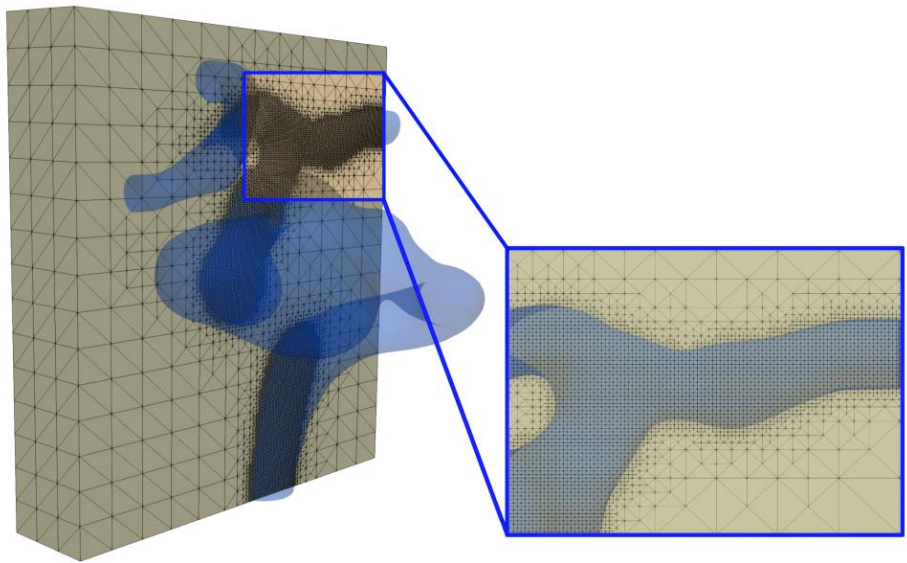

**Figure 10.** Internal cerebral artery (ICA) case Cartesian grid with four levels of local refinement.

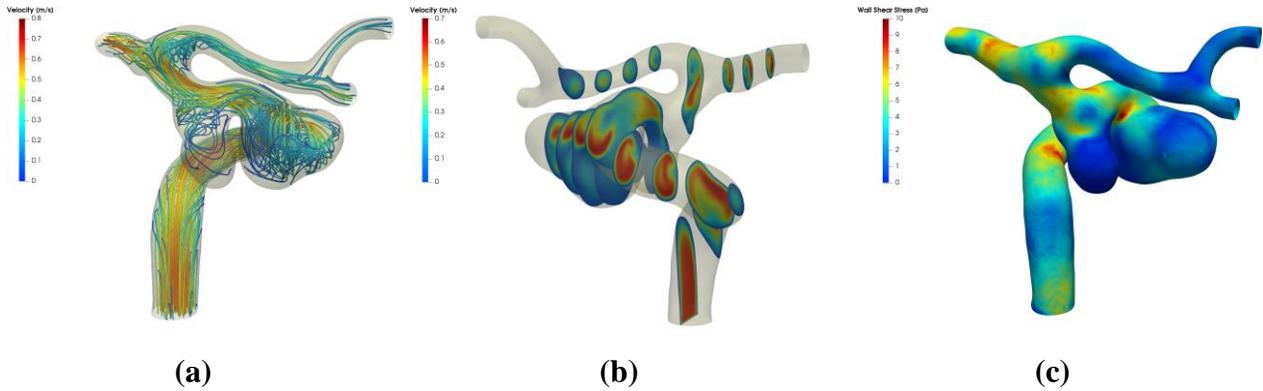

**Figure 11.** **(a)** Streamlines **(b)** velocity contours and **(c)** wall shear stress magnitude distribution (units in Pa) for blood flow through the ICA case aneurysm.

Fig. 11a shows the streamlines at the peak systole pressure. We observe that inside the larger aneurysmal dome a slow recirculation area is generated in the absence of an impinging jet. Inside the smaller aneurysmal dome, a more disturbed flow is evident leading to swirling flow patterns. The WSS magnitude on the lumen wall at the peak systolic pressure, computed via the post-processing method described previously. As expected, low WSS occurs on the aneurysmal dome, while high on the aneurysmal neck and at less extent on the walls of the parental artery (Fig. 11c). The flow pattern occurs due to the high curvature of the parent vessel which does not allow the formation of strong jets inside the aneurysmal dome. These findings agree with previous computational studies from Valen-Sendstad et al. [11] and Cebral et. al [43].

As a final flow example, we model a BAS aneurysm (Fig. 3d). We create a Cartesian grid to immerse the BA geometry, discretised using linear tetrahedral elements.

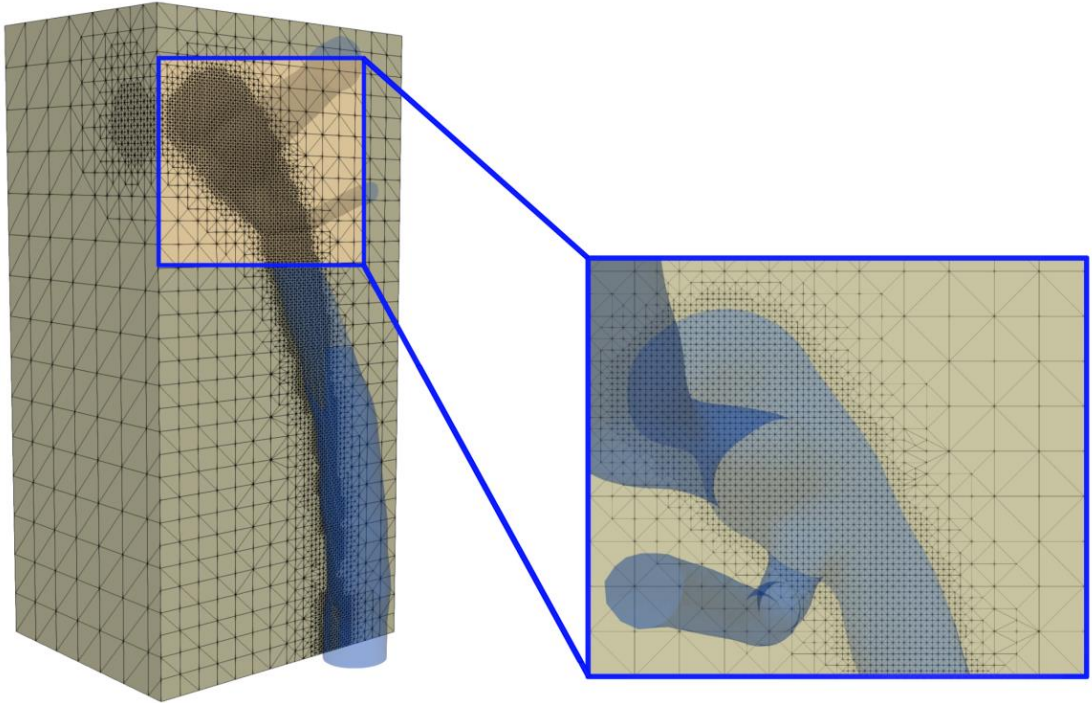

**Figure 12.** Basilar artery (BAS) aneurysm case Cartesian grid with four levels of local refinement.

We apply four levels of local mesh refinement with the resulting Cartesian grid consists of 5,542,170 tetrahedral elements (Fig. 12) and with a minimum grid spacing settling to $7.5 \times 10^{-5} m$ while we interpolate the computed results to a visualization mesh.

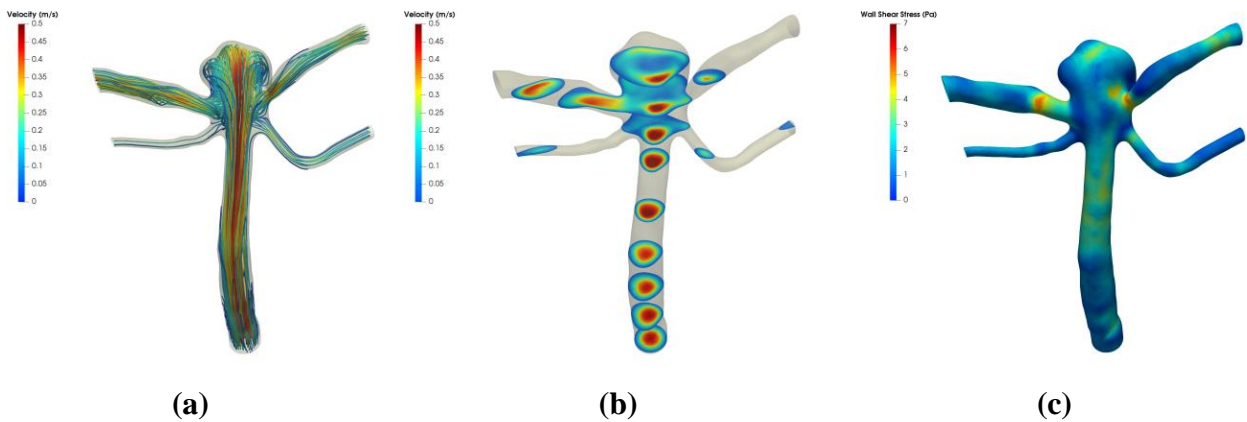

**(a)**          **(b)**          **(c)**

**Figure 13. (a)** Streamlines **(b)** velocity contours and **(c)** wall shear stress magnitude distribution (units in Pa) for blood flow through the BAS case aneurysm.

Fig. 13 illustrate the streamlines, velocity contours and the WSS at the peak systole pressure. Inside the aneurysmal dome, an impinging jet forms mainly due to the small angle between the parent artery and the aneurysm (Table 1). The WSS magnitude values are high in places where the flow impinges on the aneurysmal wall, compared to the aneurysm cases where the flow is disturbed, and re-circulations areas appear. Our findings agree with studies conducted by Sforza et al. [44] Cebral et al. [45] and Meng et al. [38] demonstrating the versatility of the present scheme.

## 5. Conclusions

In the present work, we highlight the applicability of the immersed boundary N-S solver [27] on internal flow problems with complex geometry. The numerical scheme combines the finite element (FE) method for the numerical solution of Navier-Stokes equations and the IB method to simulate fluid flow problems.

We combine Boundary Condition-Enforced IB method with the open-source finite element CFD solver Oasis to simulate blood flow in intracranial aneurysms. We validate the proposed solver considering a flow case where experimental data are available. We perform unsteady pulsatile flow computations using patient-specific cases obtained from the Aneurisk online repository. In our simulations we do not consider patient-specific inflow conditions, and therefore the wall shear stress values computed have no clinical relevance as the could possibly be over-estimated [46].

In this paper we demonstrate that our method is suitable for accurate and fast computation of the local hemodynamical field in intracranial aneurysms. The proposed scheme is simple and straightforward to apply while it requires limited to no human intervention. The burden of efficiently scaling up to a large number of patient-specific aneurysm cases has been eliminated, paving the way for future use in the clinical workflow.

**Acknowledgements**

A. Wittek and K. Miller were supported by the Australian Government through the Australian Research Council's *Discovery Projects* funding scheme (project DP160100714).

**References**

1. Kaminogo, M., M. Yonekura, and S. Shibata, *Incidence and Outcome of Multiple Intracranial Aneurysms in a Defined Population.* Stroke, 2003. **34**(1): p. 16-21.
2. Karamessini, M.T., et al., *CT angiography with three-dimensional techniques for the early diagnosis of intracranial aneurysms. Comparison with intra-arterial DSA and the surgical findings.* European Journal of Radiology, 2004. **49**(3): p. 212-223.
3. Pierot, L. and K. Wakhloo Ajay, *Endovascular Treatment of Intracranial Aneurysms.* Stroke, 2013. **44**(7): p. 2046-2054.
4. Siablis, D., et al., *Intracranial aneurysms: reproduction of the surgical view using 3D-CT angiography.* European Journal of Radiology, 2005. **55**(1): p. 92-95.
5. Cebral, J.R., et al., *Efficient pipeline for image-based patient-specific analysis of cerebral aneurysm hemodynamics: technique and sensitivity.* IEEE Transactions on Medical Imaging, 2005. **24**(4): p. 457-467.
6. Castro, M.A., et al., *Hemodynamic Patterns of Anterior Communicating Artery Aneurysms: A Possible Association with Rupture.* American Journal of Neuroradiology, 2009. **30**(2): p. 297.
7. Valencia, A., et al., *Blood flow dynamics and fluid–structure interaction in patient-specific bifurcating cerebral aneurysms.* International Journal for Numerical Methods in Fluids, 2008. **58**(10): p. 1081-1100.
8. Sforza, D.M., C.M. Putman, and J.R. Cebral, *Computational fluid dynamics in brain aneurysms.* International journal for numerical methods in biomedical engineering, 2012. **28**(6-7): p. 801-808.
9. Shojima, M., et al., *Magnitude and Role of Wall Shear Stress on Cerebral Aneurysm.* Stroke, 2004. **35**(11): p. 2500-2505.
10. Khan, M., K. Valen-Sendstad, and D. Steinman, *Narrowing the expertise gap for predicting intracranial aneurysm hemodynamics: impact of solver numerics versus mesh and time-step resolution.* American Journal of Neuroradiology, 2015. **36**(7): p. 1310-1316.
11. Valen-Sendstad, K. and D.A. Steinman, *Mind the Gap: Impact of Computational Fluid Dynamics Solution Strategy on Prediction of Intracranial Aneurysm Hemodynamics and Rupture Status Indicators.* American Journal of Neuroradiology, 2014. **35**(3): p. 536.

12. Liang, L., et al., *Towards the Clinical utility of CFD for assessment of intracranial aneurysm rupture – a systematic review and novel parameter-ranking tool.* Journal of NeuroInterventional Surgery, 2019. **11**(2): p. 153.
13. David, A.S. and M.P. Vitor, *How patient specific are patient-specific computational models of cerebral aneurysms? An overview of sources of error and variability.* Neurosurgical Focus FOC, 2019. **47**(1): p. E14.
14. Kallmes, D.F., *Identifying" truth" in computational fluid dynamics research.* AJNR-American Journal of Neuroradiology, 2011. **32**(6): p. E122.
15. Mittal, R. and G. Iaccarino, *IMMERSED BOUNDARY METHODS.* Annual Review of Fluid Mechanics, 2005. **37**(1): p. 239-261.
16. Peskin, C.S., *Flow patterns around heart valves: A numerical method.* Journal of Computational Physics, 1972. **10**(2): p. 252-271.
17. Taira, K. and T. Colonius, *The immersed boundary method: A projection approach.* Journal of Computational Physics, 2007. **225**(2): p. 2118-2137.
18. Peskin, C.S., *The immersed boundary method.* Acta Numerica, 2003. **11**: p. 479-517.
19. Gilmanov, A. and F. Sotiropoulos, *A hybrid Cartesian/immersed boundary method for simulating flows with 3D, geometrically complex, moving bodies.* Journal of Computational Physics, 2005. **207**(2): p. 457-492.
20. Mittal, R., et al., *A VERSATILE SHARP INTERFACE IMMERSED BOUNDARY METHOD FOR INCOMPRESSIBLE FLOWS WITH COMPLEX BOUNDARIES.* Journal of computational physics, 2008. **227**(10): p. 4825-4852.
21. Pinelli, A., et al., *Immersed-boundary methods for general finite-difference and finite-volume Navier–Stokes solvers.* Journal of Computational Physics, 2010. **229**(24): p. 9073-9091.
22. Colonius, T. and K. Taira, *A fast immersed boundary method using a nullspace approach and multi-domain far-field boundary conditions.* Computer Methods in Applied Mechanics and Engineering, 2008. **197**(25): p. 2131-2146.
23. Anupindi, K., et al., *A novel multiblock immersed boundary method for large eddy simulation of complex arterial hemodynamics.* Journal of Computational Physics, 2013. **254**: p. 200-218.
24. Dillard, S.I., et al., *From medical images to flow computations without user-generated meshes.* International Journal for Numerical Methods in Biomedical Engineering, 2014. **30**(10): p. 1057-1083.
25. Seo, J.-H., et al., *A Highly Automated Computational Method for Modeling of Intracranial Aneurysm Hemodynamics.* Frontiers in Physiology, 2018. **9**(681).
26. Mortensen, M. and K. Valen-Sendstad, *Oasis: A high-level/high-performance open source Navier–Stokes solver.* Computer Physics Communications, 2015. **188**: p. 177-188.
27. Bourantas, G., et al., *Immersed boundary finite element method for blood flow simulation.* arXiv preprint arXiv:2007.02082, 2020.
28. Bergersen, A.W., M. Mortensen, and K. Valen-Sendstad, *The FDA nozzle benchmark: "In theory there is no difference between theory and practice, but in practice there is".* International Journal for Numerical Methods in Biomedical Engineering, 2019. **35**(1): p. e3150.
29. Logg, A., et al., *Automated Solution of Differential Equations by the Finite Element Method The FEniCS Book.* 2016.

30. Bovendeerd, P.H.M., et al., *Steady entry flow in a curved pipe.* Journal of Fluid Mechanics, 2006. **177**: p. 233-246.
31. Plaza, A. and G.F. Carey, *Local refinement of simplicial grids based on the skeleton.* Applied Numerical Mathematics, 2000. **32**(2): p. 195-218.
32. Cebral, J.R., et al., *Characterization of Cerebral Aneurysms for Assessing Risk of Rupture By Using Patient-Specific Computational Hemodynamics Models.* American Journal of Neuroradiology, 2005. **26**(10): p. 2550.
33. Hoi, Y., et al., *Characterization of volumetric flow rate waveforms at the carotid bifurcations of older adults.* Physiological measurement, 2010. **31**(3): p. 291-302.
34. Caro, C.G., et al., *The Mechanics of the Circulation*. 2 ed. 2011, Cambridge: Cambridge University Press.
35. Khan, M.O., D.A. Steinman, and K. Valen-Sendstad, *Non-Newtonian versus numerical rheology: Practical impact of shear-thinning on the prediction of stable and unstable flows in intracranial aneurysms.* International Journal for Numerical Methods in Biomedical Engineering, 2017. **33**(7): p. e2836.
36. Voß, S., et al., *Fluid-Structure Simulations of a Ruptured Intracranial Aneurysm: Constant versus Patient-Specific Wall Thickness.* Computational and mathematical methods in medicine, 2016. **2016**: p. 9854539-9854539.
37. Philipp, B., et al., *A review on the reliability of hemodynamic modeling in intracranial aneurysms: why computational fluid dynamics alone cannot solve the equation.* Neurosurgical Focus FOC, 2019. **47**(1): p. E15.
38. Meng, H., et al., *High WSS or Low WSS? Complex Interactions of Hemodynamics with Intracranial Aneurysm Initiation, Growth, and Rupture: Toward a Unifying Hypothesis.* American Journal of Neuroradiology, 2014. **35**(7): p. 1254.
39. Jou, L.D., et al., *Wall Shear Stress on Ruptured and Unruptured Intracranial Aneurysms at the Internal Carotid Artery.* American Journal of Neuroradiology, 2008. **29**(9): p. 1761.
40. Sugiyama, S.-I., et al., *Hemodynamic Analysis of Growing Intracranial Aneurysms Arising from a Posterior Inferior Cerebellar Artery.* World Neurosurgery, 2012. **78**(5): p. 462-468.
41. Joldes, G.R., et al., *Modified moving least squares with polynomial bases for scattered data approximation.* Applied Mathematics and Computation, 2015. **266**: p. 893-902.
42. Ford, M.D., et al., *Characterization of volumetric flow rate waveforms in the normal internal carotid and vertebral arteries.* Physiological Measurement, 2005. **26**(4): p. 477-488.
43. Cebral, J.R., et al., *Analysis of hemodynamics and wall mechanics at sites of cerebral aneurysm rupture.* Journal of NeuroInterventional Surgery, 2015. **7**(7): p. 530.
44. Sforza, D.M., et al., *Hemodynamic Analysis of Intracranial Aneurysms with Moving Parent Arteries: Basilar Tip Aneurysms.* International journal for numerical methods in biomedical engineering, 2010. **26**(10): p. 1219-1227.
45. Cebral, J.R., S. Hendrickson, and C.M. Putman, *Hemodynamics in a Lethal Basilar Artery Aneurysm Just before Its Rupture.* American Journal of Neuroradiology, 2009. **30**(1): p. 95.
46. McGah, P.M., et al., *Accuracy of computational cerebral aneurysm hemodynamics using patient-specific endovascular measurements.* Annals of biomedical engineering, 2014. **42**(3): p. 503-514.